\documentclass[journal]{IEEEtran}

\usepackage{cite}
\usepackage{amsmath,lipsum}
\usepackage{amsfonts}
\usepackage{algorithm}
\usepackage{algorithmic}
\usepackage{hyperref}
\ifCLASSOPTIONcompsoc
  \usepackage[caption=false,font=normalsize,labelfont=sf,textfont=sf]{subfig}
\else

 \usepackage[caption=false,font=footnotesize]{subfig}
\fi
\usepackage{float}
\usepackage{graphicx}
\usepackage{stfloats}
\usepackage{makecell}
\usepackage{color}
\usepackage{svg}
\usepackage[justification=centering]{caption}

\hypersetup{hypertex=true,
colorlinks=true,
linkcolor=blue,
anchorcolor=blue,
citecolor=blue}

\newtheorem{lemma}{Lemma}

\begin{document}

\title{Near-Field Channel Estimation for Extremely
Large-Scale Array Communications: A model-based deep learning approach}
\author{Xiangyu Zhang,~\IEEEmembership{Student Member,~IEEE,}, Zening Wang, Haiyang Zhang,~\IEEEmembership{Member,~IEEE,}\\
        and Luxi Yang,~\IEEEmembership{Senior Member,~IEEE}
\thanks{X. Zhang, and L. Yang are with the School of Information Science and Engineering, and the National Mobile Communications Research Laboratory, Southeast University, Nanjing 210096, China, and also with the Pervasive Communications Center, Purple Mountain Laboratories, Nanjing, China (e-mail: xyzhang@seu.edu.cn;  lxyang@seu.edu.cn);  Z. Wang is with the AsiaInfo Technologies Limited, Beijing, China (e-mail:zeningw2022@126.com); H. Zhang is with the School of Communication and Information Engineering, Nanjing University of Posts and Telecommunications, Nanjing, China. (e-mail: 20220142@njupt.edu.cn).}
% Z. Wang and H. Zhang are with the School of Communication and Information Engineering, Nanjing University of Posts and Telecommunications, Nanjing, China. (e-mail:zeningw2022@126.com; 20220142@njupt.edu.cn).}
}

\maketitle
\begin{abstract} 
Extremely large-scale massive MIMO (XL-MIMO) has been reviewed as a promising technology for future wireless communications. The deployment of XL-MIMO, especially at high-frequency bands, leads to users being located in the near-field region instead of the conventional far-field. This letter proposes efficient model-based deep learning algorithms for estimating the near-field wireless channel of XL-MIMO communications. In particular, we first formulate the XL-MIMO near-field channel estimation task as a compressed sensing problem using the spatial gridding-based sparsifying dictionary, and then solve the resulting problem by applying the Learning Iterative Shrinkage and Thresholding Algorithm (LISTA). Due to the near-field characteristic, the spatial gridding-based sparsifying dictionary may result in low channel estimation accuracy and heavy computational burden. To address this issue, we further propose a new sparsifying dictionary learning-LISTA (SDL-LISTA) algorithm that formulates the sparsifying dictionary as a neural network layer and embeds it into LISTA neural network. The numerical results show that our proposed algorithms outperform non-learning benchmark schemes, and SDL-LISTA achieves better performance than LISTA with ten times atoms reduction.
\end{abstract}


\IEEEpeerreviewmaketitle

\section{Introduction} 
Extremely large-scale MIMO (XL-MIMO) has been viewed as a promising physical technology for future sixth-generation (6G) communications \cite{bjornson2019massive}. The main difference between XL-MIMO systems and the traditional multiple-antenna communication system is that the user and scatterers likely locate in the near field region of the XL-MIMO transmitter, where the electronic waves follow the spherical wave model instead of the far-field planner model \cite{dongNearFieldSpatialCorrelation2022a}. This shift brings new opportunities to benefit wireless communications, such as the beam-focusing effect, which enables improving the transmission rate of near-field  multiuser communications \cite{9738442}, as well as enhancing the energy transfer efficiency of near-field wireless power transfer systems \cite{zhang2022near}.  

The benefit of near-field beam focusing relies on accurate channel estimation, which is an important but quite challenging task. This is because, 
% the benefits brought by the XL-MIMO are accompanied by challenges, especially in channel estimation. Firstly, the hybrid antenna array architecture is the most efficient way to construct XL-MIMO. Compared to the traditional communication systems, the measurement compressed ratio of hybrid antennae, i.e. the ratio between the number of antennae and the number of RF chains, is much bigger. Then, estimating the uncompressed channel from the compressed receive signal is much harder for XL-MIMO. On the other side, 
the near-field spherical-wave channel model is more complicated than the far-field plane-wave channel model. For example, the near-field channel model incorporates not only incident angles but also distances between the BS and the user or scatter. So far, very limited literature has investigated the near-field channel estimation problem \cite{9693928, hanChannelEstimationExtremely2020}. In \cite{9693928}, a hybrid near and far field channel is formulated and represented in the polar domain. As the channel can be treated as a spatial sparse signal, the authors in  \cite{9693928} apply the orthogonal matching pursuit (OMP) to estimate the hybrid-field channel parameters. In \cite{hanChannelEstimationExtremely2020}, the authors consider a non-stationary near-field channel model and estimate the near-field channel with the same algorithm as \cite{9693928}. 

The existing XL-MIMO channel estimation methods are the same as the algorithms for the conventional far-field hybrid antenna array, i.e., introducing the channel sparse representation method and applying the compressed sensing (CS) algorithms such as OMP to estimate channel parameters \cite{7458188}. The common sparse representation method is spatial gridding, which partitions the spacing into the grid and assumes the signal incidents from these grids. In general, due to the limited granularity of grids, this assumption limits the upper boundary of channel estimation accuracy \cite{6847111}. 
Besides, this sparse representation method suffers from a huge dictionary due to the near-field space needing to be partitioned in both the angle axis and the distance axis, resulting in high computational complexity. Moreover, the sparsifying dictionary generated by spatial gridding may not satisfy the restricted isometry property because the array response vector in the distance axis shows a high correlation. In this case, the recovery accuracy of CS algorithms will be unavoidably degraded \cite{eldar2012compressed}.

In this letter, we investigate the near-field channel estimation problem for multi-user XL-MIMO communication systems. The XL-MIMO base station (BS) estimates the wireless channels between itself and multiple signal-antenna users using the collected orthogonal pilot signals sent by users. We consider the communication system operating in high-frequency bands. Thus, the wireless channel can be assumed to be sparse. For such a scenario, we propose model-based deep learning approaches to solve the channel estimation problem with high accuracy and efficiency. 
In particular, the main contributions of this letter are summarized as follows:
\begin{itemize}
    \item We first formulate the near-field channel estimation as a CS problem using the spatial gridding-based sparsifying dictionary, which allows for estimating the sparse channel from a small number of observed pilot signals. Then, we apply the model-based deep learning approach-Learning Iterative Shrinkage and Threshold Algorithm (LISTA),  to solve the corresponding sparse parameter estimation problem efficiently.
    \item Next, we prove that the LISTA is limited to achieve high channel estimation accuracy and suffers high computational complexity raised by the spatial gridding dictionary. A direct way is to optimize the sparsifying dictionary. We propose Sparsifying Dictionary Learning-LISTA (SDL-LISTA), which formulates the sparsifying dictionary as a neural network layer and embeds it inside each layer of LISTA. Thus, a sparsifying dictionary and the parameter of LISTA can be trained when LISTA training. The learned sparsifying dictionary reaches high accuracy because it fits on the channel distribution. 
    \item We provide numerical examples to validate the effectiveness of our proposed algorithms. In particular, we show that our proposed algorithms outperform non-learning benchmark schemes, including OMP and Fast ISTA (FISTA), and SDL-LISTA achieves better performance than LISTA with ten times atoms reduction.
\end{itemize}

The rest of the letter is organized as follows. In Section II, we introduce the considered XL-MIMO systems and the formulated near-field channel estimation problem. In Section III, we present our algorithms, i.e., LISTA and SDL-LISTA. Simulation results and conclusions are provided in Section IV and Section V, respectively.

\section{System Model}

\begin{figure}[h]
\centering
\includegraphics[width=0.45\textwidth]{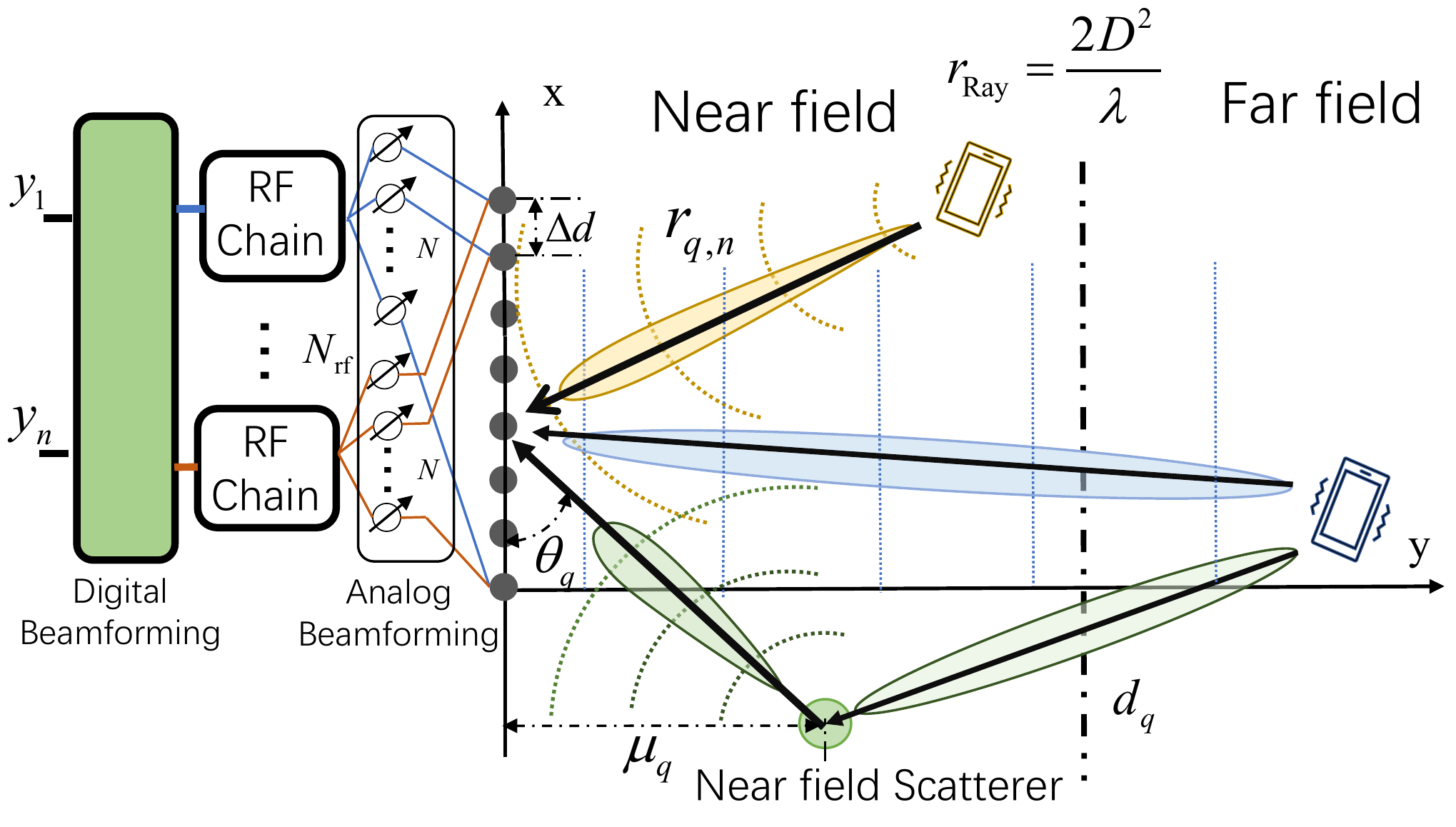}
\caption{The system layout of extremely large arrays and the signal propagation paths}
\label{fig:system layout}
\end{figure}

As shown in Fig.~\ref{fig:system layout}, we consider a multi-user XL-MIMO communication system consisting of an XL-array BS and $M$ signal-antenna users, operating at high-frequency bands. The users are located in the near-field region of XL-array BS. We consider the XL-MIMO BS adopts a hybrid antenna architecture, where $N$ antenna elements are connected to $N_\text{RF}$ RF chains through a phase shifter-based analog combiner network \cite{6847111}.  The number of RF chains is less than the number of antenna numbers, i.e., $N_\text{RF}<N$. For such a scenario, we study the uplink near-field channel estimation problem. The users send  mutual orthogonal pilots to the BS, and the BS estimates the near-field channels according to the processed pilot signals of the phase shifter-based analog combiner. Since users send
mutual orthogonal pilots, the channel between BS and each user can be estimated individually. 
In addition, we denote the $ \lambda, c, \Delta d=\frac{\lambda}{2}$ and $D = N \Delta d$ as the wavelength, the speed of light, antenna spacing, and antenna aperture, respectively. 
\subsection{Near-Field Region and Channel Model}
A typical signal propagation area includes two regions: near-field (Fresnel area) and far-field (Fraunhofer area), which are separated by the classical Rayleigh distance $r_{\text {Ray}}=\frac{2D^{2}}{\lambda}$. Due to the utilization of the larger antenna aperture and the higher communication frequency, the near-field area will significantly expand. Thus users/scatterers will likely be located within the near-field region. 
As in \cite{dongNearFieldSpatialCorrelation2022a, luNearFieldModellingPerformance2021, 9693928 }, we consider a scattering environment in which the $Q$ scatterers are located either in the near field or far field. The received signal is assumed as the signal incident from several direct and scattering paths. Based on the bistatic radar equation, the channel between the specific user and the $n$-th antenna element of BS can be formulated as
\begin{equation}
\label{eq-channel_single_form}
\begin{aligned}
    h_n=& \sum_{q=1}^Q \frac{\lambda \sqrt{\rho}}{(4 \pi)^{3 / 2}r_{q, n}} e^{-j kr_{q, n}}\\
    % \mathop  \simeq \limits^{(a)}  & \sum_{q=1}^Q \frac{\lambda \sqrt{\sigma_q}}{(4 \pi)^{3 / 2}r} e^{-j k (d_{q}+\mu_q) +j \psi_q} e^{-j k (\frac{(n\Delta d)^2}{2\mu_q} - n\Delta d \sin \theta_q)}
\end{aligned},
\end{equation}
where $r_{q, n}$ is the length of the propagation path. $\beta_{q, n} =  \frac{\lambda}{(4 \pi)^{3 / 2}r_{q, n}}$ represents the free space path loss. $\rho$ represents the scattering cross sections. For the direct path, $\rho=1$. $k = 2\pi/\lambda$. 

Denote $\mathbf{h}  = \left[h_0, \cdots, h_{N-1}\right]^\text{T}$ as the near-field channel vector. We simplify the channel $\mathbf{h}$ by approximating the distance as $ r_{q, n} \simeq  d_{q}+\mu_q + \frac{(n\Delta d)^2}{2\mu_q} - n\Delta d \sin \theta_q$, where $\mu_q$ and $\theta_q$ is the y-axis coordinate, and the incident angle of the  last-hop scatter of the path $q$\cite{4155681}. Meanwhile, we ignore the difference of path loss on the different antennas, i.e., setting $\beta_{q} = \beta_{q, 0} = \cdots = \beta_{q, N}$. Then, $\mathbf{h}$ can be represented by 
\begin{equation}
\label{eq-channel}
    \mathbf{h} = \mathbf{A}_Q\boldsymbol{\alpha}_Q ,
\end{equation}
where $\mathbf{A}_Q\in \mathbb{C}^{N \times Q}$, and the $q$-th column vector $\mathbf{a}_{q}$ is the array response vector of 
$q$-th path, given by 
\begin{equation}
    \mathbf{a}_{q}(\mu_q, \theta_q ) = [\cdots, e^{-j k\left(\frac{(n\Delta d)^2}{2\mu_q} - n\Delta d \sin \theta_q \right)}, \cdots]^{\mathrm{T}},
    \label{eq-array responds vector}
\end{equation}
$\boldsymbol{\alpha}_Q$ represents the path loss, whose elements is $\alpha_q = \beta_q e^{-jk(\mu_q+d_{q})}$.

Equation \eqref{eq-channel} implies that each element of $\mathbf{h}$ depends only on a few angles and distances, which enables the spatial sparse representation given by 
\begin{equation}
 \label{eq-sparse_representation}
   \mathbf{h} \simeq  \mathbf{A}\boldsymbol{\alpha}, 
\end{equation}
where $\mathbf{A}\in \mathbb{C}^{N \times G}$ is the sparsifying dictionary and $G$ is the number of atoms. As in \cite{9693928}, $\mathbf{A}$ is constructed by spatial gridding, which discretizes the spaces into $G_a$ partitions in $\mu$ axis and $G_d$ partitions in $\psi$ axis. Namely, the space is partitioned in $G =G_a \times G_d$ two-dimensional grids. Then, $\mathbf{A}$ can be formulated as 
 \begin{equation}
 \label{eq:spatial}
     \mathbf{A} = [\mathbf{a}(\mu_1, \theta_1 ),\dots,\mathbf{a}(\mu_G, \theta_G )].
 \end{equation}
 $\boldsymbol{\alpha}\in \mathbb{C}^{G \times 1}$ is the sparse channel gain corresponding to $\mathbf{A}$. When there is no channel path in $g$-th grid, the $g$-th element of $\boldsymbol{\alpha}$, denoted by $\alpha_g$ is equal to zero, i.e., $\alpha_g = 0$. As $Q \ll N \ll G$
 $\boldsymbol{\alpha}$ is highly sparse.

\subsection{Problem Formulation}

\label{subsec:problem}

The observed pilot signal of the BS over $T$ pilot slots, denoted by $\mathbf{Y} \in \mathbb{C}^{N_{\rm{RF}} \times T}$, is given by
\begin{equation}
   \label{received_signal1}
\begin{aligned}
\mathbf{Y}= \mathbf{W}\mathbf{h} \mathbf{s} +  \mathbf{W}{\mathbf{n}}
\end{aligned}
\end{equation}
where $\mathbf{W}\in \mathbb{C}^{N_{\rm{RF}} \times N}$ is the configurable analog combiner matrix of the hybrid antenna; $\mathbf{s}\in \mathbb{C}^{1 \times T}$ denotes the orthogonal pilot sequence of the specific user, satisfying $\mathbf{s}{\mathbf{s}}^{\rm{H}}=1$; ${\mathbf{n}} \in  \mathbb{C}^{N \times T}$ represents the noise matrix with each element following the distribution $\mathcal{N}(0,\sigma^2)$. As the orthogonal pilot is known at the BS side, we have
\begin{equation}
 \label{received_signal2}
\mathbf{y} = \mathbf{Y}{\mathbf{s}}^{\rm{H}} = \mathbf{W}\mathbf{h} +  \mathbf{W}{\mathbf{n}}\mathbf{s}^{\rm{H}}\simeq  \mathbf{W}\mathbf{A}\boldsymbol{\alpha} +  \hat{{\mathbf{n}}} ,
\end{equation}
where $\hat{{\mathbf{n}}}=\mathbf{W}{\mathbf{n}}\mathbf{s}^{\rm{H}}$ is the equivalent noise.

Our objective is to estimate the near-field channel $\mathbf{h}$ from $\mathbf{y}$, which is an underdetermined problem because of $N_{\rm{RF}} \ll N$. To tackle this challenge, we utilize the sparse representation in \eqref{eq-sparse_representation} that turns the estimation of channel $\mathbf{h}$ to estimate the sparse signal $\boldsymbol{\alpha}$ equivalently. The estimation of $\boldsymbol{\alpha}$ is a typical CS problem \cite{eldar2012compressed}, which is formulated as, 
\begin{equation}
\begin{aligned}
\label{LASSO}
 ~  \min_{\boldsymbol{\alpha}} ~&\|\mathbf{y} -   \mathbf{W}\mathbf{A}\boldsymbol{\alpha}  \|_2  \\
\text { s.t }~ &\|\boldsymbol{\alpha}\|_{0} < \epsilon \\
\end{aligned},
\end{equation}
where $\epsilon$ is a given threshold parameter related to $\sigma^2$.

Problem \eqref{LASSO} is non-convex due to the non-convex $\ell_0$ norm constraint. In the next section, we propose two model-based deep learning algorithms to solve it efficiently.

\section{Proposed Algorithm}
\subsection{LISTA-based Near Field Channel Estimation}
\label{subsec:LISTA}

By relaxing the $\ell_0$ norm constraint to $\ell_1$ norm constraint, we transfer \eqref{LASSO} into a tractable form, given by
\begin{equation}
\begin{aligned}
\label{eq-lasso}
 ~  \min_{\boldsymbol{\alpha}} ~&\|\mathbf{y} -   \mathbf{\Psi}\boldsymbol{\alpha}  \|_2  + \xi \|\boldsymbol{\alpha}\|_{1}\\
\end{aligned},
\end{equation}
where $\mathbf{\Psi} = \mathbf{W}\mathbf{A}$ and $\xi$ is a regularization parameter.

Problems \eqref{LASSO} and \eqref{eq-lasso} share the same solution when the sensing matrix $\mathbf{\Psi}$ has the restricted isometry property (RIP) \cite{eldar2012compressed}.
Problem \eqref{eq-lasso} is a typical sparse recovery problem.  One of the commonly used approaches to solve \eqref{eq-lasso} is the Iterative Shrinkage and Thresholding Algorithm (ISTA) \cite{4959678}, which recovers the sparse $\boldsymbol{\alpha}$ by iterating the following recursive equation: 
\begin{equation}
\label{ISTA_eq}
\boldsymbol{\alpha}^{(t+1)} \!=\! h_{(\eta)}\left((I - \frac{1}{\lambda_{\max}} \mathbf{\Psi}^{\rm{T}} \mathbf{\Psi})\boldsymbol{\alpha}^{(t)}
+\frac{1}{\lambda_{\max}} \mathbf{\Psi}^{\rm{T}}\mathbf{y}\right),
\end{equation}
where $h_{(\eta)}$ is the element-wise soft shrinkage function, given by $[h_{(\eta)}]_g = \rm{sign}([\boldsymbol{\alpha}]_g)(|[\boldsymbol{\alpha}]_g|-\eta)_{+}$, where $[\cdot]_g$ is the $g$-th elements of vector; $\rm{sign}(\cdot)$ returns the sign of a scalar; $(\cdot)_{+}$ means $\max(\cdot,0)$; $\eta$ is a constant threshold.
$\lambda_{\max}$ is the maximum eigenvalue of $\mathbf{\Psi}^{\rm{T}}\mathbf{\Psi}$. $\mathbf{I}$ is the identity matrix.

% , the neural network of LISTA receives $\mathbf{y}$ and $\boldsymbol{\alpha}^{(0)}$, the initiate value of $\boldsymbol{\alpha}$, and outputs the recovered $\boldsymbol{\alpha}$. 

\begin{figure}[!htb]
\centering
\includegraphics[height=1.4in]{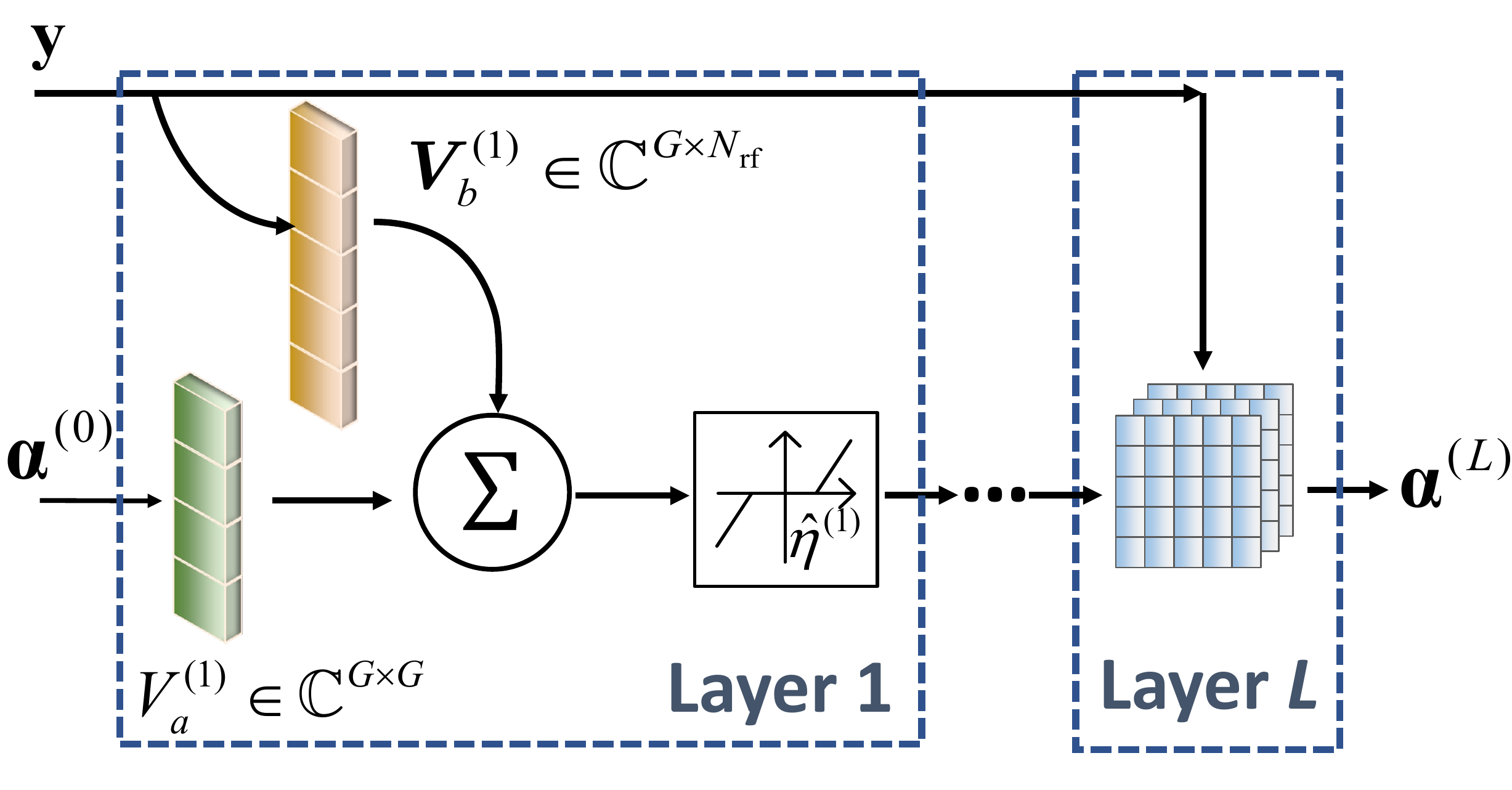}
\caption{The block diagram of LISTA algorithm}
\label{fig:LISTA}
\end{figure}

The recovering accuracy of ISTA is heavily influenced by $\eta$ and iteration time, which are hard to select for each $\boldsymbol{\alpha}$. To achieve higher estimation accuracy and efficiency for \eqref{eq-lasso}, we apply LISTA, a neural network version of the ISTA, which overcomes the disadvantages of ISTA by fixing the iteration time and learning $\eta$ from the data \cite{gregorLearningFastApproximationsc}.
The key idea of LISTA is to unfold each iteration of ISTA to a neural network layer. The input and output relationship of the $l$-th layer is formulated as
\begin{equation}
\label{LISTAequation2}
\boldsymbol{\alpha}^{(l+1)}=h_{(\hat{\eta}^{(l)})}\left(\mathbf{V}_a^{(l)}\boldsymbol{\alpha}^{(l)}+\mathbf{V}_b^{(l)}\mathbf{y}\right),
\end{equation}
where $\mathbf{V}_a$, $\mathbf{V}_b$, and $\hat{\eta}$ are learnable parameters for representing $I - \frac{1}{\lambda_{\max}} \mathbf{\Psi}^{\rm{T}} \mathbf{\Psi}$, $\frac{1}{\lambda_{\max}} \mathbf{\Psi}^{\rm{T}}$ and $\eta$, respectively.
% We represent LISTA as $\boldsymbol{\alpha}^{(L)} =  \mathcal{N} ({\bf{y}}_s| \mathbf{V}_a, \mathbf{V}_b, \hat{\eta})$, where $\mathcal{N}$ is the neural network. 

Fig.~\ref{fig:LISTA} illustrates the neural network of LISTA with $L$ layers. The output feature of LISTA network $\boldsymbol{\alpha}^{(L)}$ denotes the solution of \eqref{eq-lasso}. Then, the near-field channel $\mathbf{h}$ can be obtained from \eqref{eq-sparse_representation} accordingly.
As a data-driven method, the parameters of LISTA are updated by minimizing the loss function 
$
    \mathcal{L} = \sum_{s=1}^S\left\| \boldsymbol{\alpha}^*_s - \boldsymbol{\alpha}^{(L)} \right\|_2,
    \label{eq-lossfunction}
$
where $\boldsymbol{\alpha}^*_s$ is the accurate channel gain. We denote $\mathbf{y}_s$ as the corresponding received signal of $\boldsymbol{\alpha}^*_s$. Then, the data set can be formulated as $\mathbb{D} = \left\{[\mathbf{y}_s, \boldsymbol{\alpha}^*_s]| \mathbf{y}_s = \mathbf{\Psi}\boldsymbol{\alpha}^*_s + \mathbf{n}, s = 1,..., S \right\}$; $S$ is the total number of samples.

\subsection{SDL-LISTA-based Near Field Channel Estimation}

The above Section \ref{subsec:LISTA} provides a  LISTA-based approach to 
study the new near-field channel estimation problem. It is worth noting that in Section \ref{subsec:LISTA}, the sparsifying dictionary $\bf A$ is constructed by simple spatial gridding. For the task of near-field channel estimation, the fixed spatial gridding $\bf A$ brings several challenges, which will be detailed later.
\begin{lemma}[\cite{eldar2012compressed}]
\label{lemma1}
    To accurately recover $\boldsymbol{\alpha}$ from $\mathbf{y}$ in \eqref{eq-lasso}, $\mathbf{\Psi}$ should satisfy the RIP condition, i.e., $\nu(\mathbf{\Psi})<\frac{1}{2Q-1}$, where $\nu(\mathbf{\Psi})$ is defined as
    \begin{equation}
            \nu(\mathbf{\Psi}) = \max _{1 \leq t<v \leq G} \frac{\left|\left\langle \boldsymbol{\psi}_t(\mu_t, \theta_t ), \boldsymbol{\psi}_v(\mu_v, \theta_v )\right\rangle\right|}{\left\|\boldsymbol{\psi}_t(\mu_t, \theta_t )\right\|_2\left\|\boldsymbol{\psi}_v(\mu_v, \theta_v )\right\|_2},
    \end{equation}
\end{lemma}
where $\boldsymbol{\psi}_t$ and $\boldsymbol{\psi}_v$ are any two column of $\mathbf{\Psi}$.
\begin{lemma}
\label{lemma2}
The spatial gridding dictionary $\mathbf{A}$ defined in \eqref{eq:spatial} may not support $\mathbf{\Psi}$ to guarantee Lemma \ref{lemma1}.
\end{lemma}
\begin{IEEEproof}
Please refer to Appendix \ref{appendix1}.
\end{IEEEproof}

Lemmas \ref{lemma1} and \ref{lemma2} indicate that the spatial gridding dictionary $\mathbf{A}$ may degrade the estimation accuracy of near-field channels. Meanwhile, the size of $\mathbf{A}$ is much larger than that of far-field spatial gridding dictionary, resulting in serious storage and computational burden. 
% In particular, the high computational burden may lead to the LISTA failure to recover $\boldsymbol{\alpha}$ in a channel correlation time.   

% \begin{remark}
% \label{remark-timelimited}
%      As users are located in the near-field, the spatial gridding dictionary for XL-MIMO systems often has more atoms than the far-field spatial gridding dictionary. A calculation burden is raised by the matrix multiplication of $\mathbf{A}$ in \eqref{ISTA_eq} and \eqref{LISTAequation2}, which may lead to the ISTA and LISTA failure to recover $\boldsymbol{\alpha}$ in channel correlation time.   
% \end{remark}

% Comment \ref{remark-timelimited} explains the calculation inefficiency brought by $\mathbf{A}$. 

\begin{figure}[!htb]
\centering
{\includegraphics[width=0.5\textwidth]{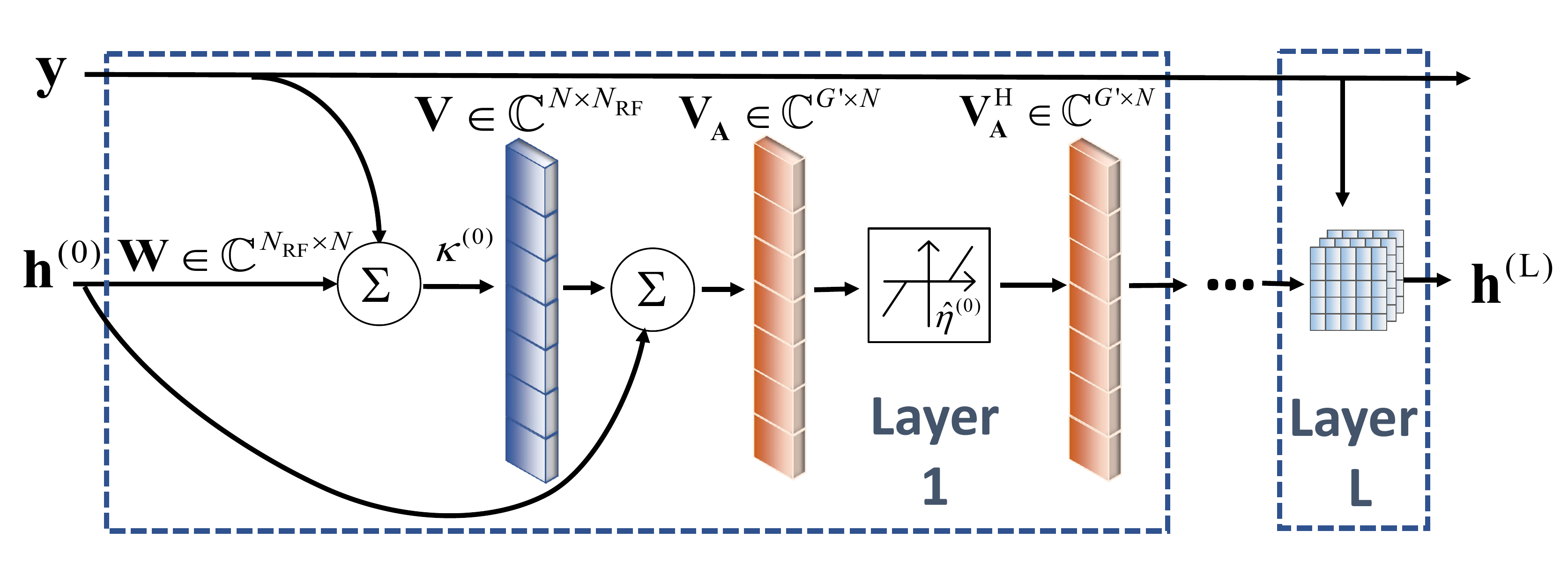}}
\caption{The block diagram of SDL-LISTA}
\label{fig:SDL-LISTA}
\end{figure}

To address these challenges, we propose a novel method, SDL-LISTA, that learns a sparsifying dictionary from data to achieve high channel estimation accuracy with reduced computational complexity by utilizing fewer atoms. The sparsifying dictionary optimization problem is formulated as 
\begin{equation}
\begin{aligned}
\min_{\mathbf{A}} ~&{\mathbb{E}}\left\{ \left\| {\mathbf{h}}^*- \mathbf{A} \mathcal{N} ({\bf{y}}_s| \mathbf{V}_a^*, \mathbf{V}_b^*, \hat{\eta}^*)  \right\|_2 \right\},
\end{aligned}
\label{eq-statisticaloptimizationQA}
\end{equation}
where $\mathcal{N} ({\bf{y}}_s| \mathbf{V}_a, \mathbf{V}_b, \hat{\eta})$ represents the LISTA neural network. $\mathbf{h}^*$ represents the noise-free channel, wherein the incident angle and distance of channel path follow a distribution \cite{6829967}. $\mathbf{V}_a^*$ and $\mathbf{V}_b^*$ and $\hat{\eta}^*$ are well-trained parameters under a given $\mathbf{A}$.

Directly solving \eqref{eq-statisticaloptimizationQA} with alternating iteration is challenging because it needs to train the LISTA whenever $\mathbf{A}$ is updated. We turn this problem into a model-based learning problem that trains a desired $\mathbf{A}$ and the parameters of LISTA jointly. To realize this, we map $\mathbf{A}$ as a learnable parameter $\mathbf{V}_{\mathbf{A}} \in \mathbb{C}^{N\times G'}$ and construct it as a sparsifying dictionary learning layer, where $G'$ is the number of atoms. Then, we embed this layer into the LISTA and propose the SDL-LISTA, which can be formulated as 
\begin{equation}
\label{ALISTAequation}
\mathbf{h}^{(l+1)} \!=\! \mathbf{V}_{\mathbf{A}}^{{\rm{H}}} \cdot h_{(\hat{\eta}^{(l)})}\left(\mathbf{V}_{\mathbf{A}}\left(
\mathbf{h}^{(l)} \!-\! \kappa^{(l)}\mathbf{V}(\mathbf{W}\mathbf{h}^{(l)} \!-\!\mathbf{y}\right)\right).
\end{equation}

In \eqref{ALISTAequation}, we adopt another form of LISTA proposed in \cite{chen_theoretical_2018}, in which $\mathbf{V}\in \mathbb{C}^{N \times N_{\rm{RF}}}$, $\hat{\eta}$ and $\kappa$ are learnable parameters. $\mathbf{V}$ replaces the learnable parameters $\mathbf{V}_a$ and $\mathbf{V}_b$ used in \eqref{LISTAequation2}, and it is shared among all layers. Meanwhile, it should be noticed that the sparsifying dictionary learning layer $\mathbf{V}_{\mathbf{A}}$ and $\mathbf{V}_{\mathbf{A}}^{{\rm{H}}}$ are embedded before and after the soft shrink function, respectively. $\mathbf{V}_{\mathbf{A}}$ transforms $\mathbf{h}$ into sparse form and $\mathbf{V}_{\mathbf{A}}^{{\rm{H}}}$ transforms it back.

The block diagram of SDL-LISTA is illustrated in Fig.~\ref{fig:SDL-LISTA}. Compared with LISTA, SDL-LISTA directly outputs the channel $\mathbf{h}$ instead of $\boldsymbol{\alpha}$. Thus, the neural network is trained using  the data set $\mathbb{D}_{{\text{SDL}}}  = \left\{[\mathbf{y}_s, \mathbf{h}^*_s]| \mathbf{y}_s =  \mathbf{W} \mathbf{h}^*_s + \mathbf{n}, s = 1,..., S \right\}$. The loss function is given by
$
    \mathcal{L}_{{\text{SDL}}} = \sum_{s=1}^S\left\| \mathbf{h}^*_s - \mathbf{h}^{(L)} \right\|_2.
    \label{eq-lossfunction}
$
The SDL-LISTA-based near-field channel estimation approach is summarized in Algorithm \ref{alg2}.

% \begin{figure}[!htb]
% \centering
% {\includegraphics[width=0.5\textwidth]{systemmodel/MBL.pdf}}
% \caption{The block diagram of SDL-LISTA}
% \label{fig:SDL-LISTA}
% \end{figure}

\begin{algorithm}[h]
\caption{SDL-LISTA for near-field channel estimation}
\label{alg2}
\begin{algorithmic}[1]
\STATE Initialize the parameters $\mathbf{V}$, $\mathbf{V}_\mathbf{A}$, $\hat{\eta}$ and $\kappa$. Collect the data set $\mathbb{D}_{{\text{SDL}}}  = \left\{[\mathbf{y}_s, \mathbf{h}^*_s]| \mathbf{y}_s =  \mathbf{W} \mathbf{h}^*_s + \mathbf{n}, s = 1,..., S \right\}$. 
\STATE \textbf{While} 
\STATE \quad Sample a batch of data from the data set $\mathbb{D}_{{\text{SDL}}}$;
\STATE \quad Calculate the output of neural network $\mathcal{N} (\bf{y})$ and calculate the loss $\mathcal{L}_{{\text{SDL}}}$;
\STATE \quad Update the parameters $\mathbf{V}$, $\mathbf{V}_\mathbf{A}$, $\hat{\eta}$ and $\kappa$ using backpropagation. 
\STATE \textbf{Until} $\mathcal{L}_{{\text{SDL}}}$ converge. 
\STATE \textbf{Output}: A well-trained neural network $\mathcal{N}_{{\text{SDL}}}$.
\end{algorithmic}
\end{algorithm}

% \begin{remark}
% \label{remark:SDL-LISTA}
% The problem \eqref{eq-statisticaloptimizationQA} finds a basis for the channel. The number of atoms required for the basis depends only on the distribution of incident angle and distance and the error upper boundary. It is more likely to find a basis for channels that utilize much fewer atoms than the antenna number $N$. 
% \end{remark}

% \begin{remark}
% \label{remark:LISTA-SDO}

% \end{remark}

SDL-LISTA is expected to achieve better channel estimation performance than LISTA. This is because the loss function for SDL-LISTA is directly calculated from the channel instead of the sparse signal, inherently eliminating the sparse representation error.

% Meanwhile, we replace the expectation in \eqref{eq-statisticaloptimizationQA} with the sample from the data set $\mathbb{D}_{SDL}  = \left\{[\mathbf{y}_s, \mathbf{h}^*_s]| \mathbf{y}_s =  \mathbf{W} \mathbf{h}^*_s + \mathbf{n}, s = 1,...,S \right\}$. Then the problem \eqref{eq-lossfunction} can be formulated as a sparsifying dictionary learning problem, given by 
% \begin{equation}
% \begin{aligned}
% \min_{\mathbf{A}} ~&\underset{\{{\bf{y}}_s, {\mathbf{h}}^*\} \in \mathbb{D}_{SDL} }{\mathbb{E}}\{ \| \mathbf{h}^*- \mathbf{A} \mathcal{N} ({\bf{y}}_s| \mathbf{V}_a^*, \mathbf{V}_b^*, \hat{\eta}^*) \|_2  \},
% \end{aligned}
% \label{eq-leanringdictionary}
% \end{equation}
% where  As \eqref{eq-lossfunction} and \eqref{eq-leanringdictionary} share the same objective, we can solve the sparsifying dictionary learning problem while training LISTA. To realize it, 

\section{Numberical Result}
In this section, we provide numerical experiments to demonstrate the performance of our proposed near-field channel estimation approaches. In the following experiments, we consider a $N=128$ uniform linear array with element space $\Delta d = \lambda/2$, which works at the carrier frequency 28 GHz ($\lambda$ =  1.07 cm). The near-field distance under such a setting is around  87 m.  The channels are generated according to \eqref{eq-channel_single_form}. 
 The incident angle $\phi = \sin\theta$  follows a Gaussian mixture distribution $\mathcal{N}(\boldsymbol{\mu}_\phi, \boldsymbol{\Sigma}_\phi)$, where $\boldsymbol{\mu}_\phi=\left[-0.6, -0.45, -0.2, 0.3, 0.6\right]$ and $\boldsymbol{\Sigma}_\phi = 0.15\mathbf{I}$. $\mu_q$ follows the uniform distribution $\mathcal{U}(2, 100)$ and $d_q$ follows $\mathcal{U}(0, 100)$. $Q$ is an integer randomly selected between 2 to 6. 

We utilize the Adam optimizer with the learning rate as $1e^{-4}$ in LISTA and SDL-LISTA. The batch size is 256. The initial values of $\mathbf{V}_a, \mathbf{V}_b$ for LISTA and $\mathbf{V}, \mathbf{V}_{\mathbf{A}}$ for SDL-LISTA are initialized with uniform distribution $\mathcal{U}(0, 1)$. The initial value of $\hat{\eta}$ is $10^{-4}$. The training and test sets include 256000 and 256 samples. The SNR of the sample is within the interval [0,27] dB. We measure the channel estimation accuracy with the normalized MSE (NMSE), given by
\begin{equation}
\mathrm{NMSE}=\sum_{n=1}^N \frac{\left\|h_n-h^*_n\right\|_{2}^{2} }{\|h^*_n\|_{2}^{2}}.
\end{equation}

\begin{figure}[ht]
\centering
\includegraphics[width=0.35\textwidth]{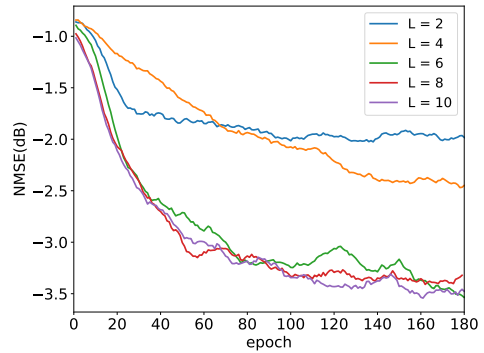}
\caption{The convergence behavior of SDL-LISTA}
\label{result:layer}
\end{figure}

In Fig.~\ref{result:layer}, we  study the convergence behavior of SDL-LISTA for different layers, considering $\mathbf{V}_\mathbf{A}$ having $G' = 256$ atoms. 
From Fig.~\ref{result:layer}, it is observed that the proposed 
SDL-LISTA can achieve fast convergence with a small number of layers. For example, SDL-LISTA has a similar estimation accuracy when $L$ is larger than 6. Also, only 100-150 epochs are needed to achieve convergence in general.
% in Fig.~\ref{result:layer}, where the NMSE of the neural networks with different layers ($L=2$, $L=4$, $L=6$, $L=8$, and $L=10$) \textit{are tested on the test set at each epoch.} All of the neural network utilized $\mathbf{V}_\mathbf{A}$ with $G' = 256$. 
% When the neural network contains more than 6 layers, convergence performance is similar. According to the results, SDL-LISTA recovers the channel in 6 iterations, which is much more efficient than FISTA, which needs hundreds of iterations.

\begin{figure}[ht]
\centering
\includegraphics[width=0.35\textwidth]{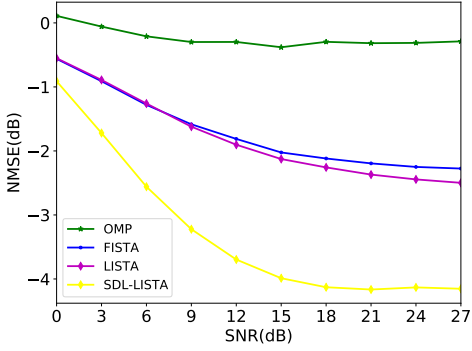}
\caption{The NMSE performance comparison}
\label{result:NMSE}
\end{figure}

In Fig.~\ref{result:NMSE}, we compare the achievable NMSE performance of our proposed model-based learning methods (LISTA and SDL-LISTA) with representative non-learning algorithms including OMP \cite{9693928} and Fast ISTA (FISTA) \cite{4959678}.
% For comparison, we also study the performance of representative non-learning algorithms including OMP \cite{9693928} and Fast ISTA (FISTA) \cite{4959678}. 
% Then, we compare the channel estimation accuracy with the samples whose SNR ranges from 0 dB to 27 dB.
The number of iterations for OMP and FISTA is set as 10 and 100, respectively. LISTA and SDL-LISTA use neural networks with 10 layers. The OMP, FISTA, and LISTA use the sparsifying dictionary with $G = 2048$ atoms generated by spatial gridding. The grids' angles are obtained by evenly partitioning $\phi \in [-1, 1]$ into 256 grids, while the grids' distances are obtained by evenly partitioning $\frac{1}{\mu_q}\in [0, 0.5]$ into 8 grids. 
From Fig.~\ref{result:NMSE}, it is observed that model-based algorithms outperform existing OMP and FISTA. For example, SDL-LISTA with $G = 256$ outperforms OMP and FISTA for 3 dB on average. Moreover, thanks to the joint training of the sparsifying dictionary, SDL-LISTA can achieve better performance than LISTA with ten times atoms
reduction.
% In addition, 
% and outperforms LISTA for 2 dB when SNR is larger than 9 dB. The NMSE of SDL-LISTA with $G = 1024$ is 4 dB less than LISTA and 6 dB more than OMP. This indicates the effectiveness of SDL-LISTA, which achieves better performance but with fewer atoms. 
% by comparing the lines of SDL-LISTA with different sizes of the sparsifying dictionary, we can observe that the NMSE of the line coincides when $G>1024$. This indicates that SDL-LISTA has a saturated upper boundary. 
% In this figure, we also show the performance difference brought by the learned sparsifying dictionaries with size $G' = 256$, $G' = 512$, $G' = 1024$, and $G' = 2048$. 

\section{Conclusion}
This letter studied the near-field channel estimation problem in XL-MIMO systems, and two model-based deep learning algorithms were proposed to estimate channel parameters efficiently. In particular,  LISTA was first applied to the near-field channel estimation problem using the spatial gridding-based sparsifying dictionary. Then, SDL-LISTA, which embeds the sparsifying dictionary into LISTA neural network, was  proposed further to enhance the near-field channel estimation accuracy with reduced complexity. Finally, simulation results were provided to verify the effectiveness of our proposed algorithms.

\appendices
\section{Proof of Lemma \ref{lemma2}}
\label{appendix1}
First, the coherence of $\boldsymbol{\Psi}$ can be rewritten as $\left|\left\langle \boldsymbol{\psi}_t(\mu_t, \theta_t ), \boldsymbol{\psi}_v(\mu_v, \theta_v )\right\rangle\right| = \left|\mathbf{a}_v(\mu_v, \theta_v)^{\rm{H}}\mathbf{W}^{\rm{H}}\mathbf{W}\mathbf{a}_t(\mu_t, \theta_t )\right|$. Without loss of generality, we assume $G=N$ and the coherence of $\mathbf{W}$ reached the Welch bound, which means $\mathbf{W}^{\rm{H}}\mathbf{W} = \mathbf{I}$ \cite{eldar2012compressed}. Then, we have 
\begin{equation}
\label{eq-appendix}
\begin{aligned}
    \nu(\mathbf{\Psi}) = &\max _{1 \leq t<v \leq G} \frac{\left|\left\langle \mathbf{a}_t, \mathbf{a}_v\right\rangle\right|}{\left\|\mathbf{a}_t\right\|_2\left\|\mathbf{a}_v\right\|_2}\\
        \simeq & \max _{1 \leq t<v \leq G} \frac{1}{N} \left|\sum_{n = 0}^{N - 1}\exp{\left(j k n \Delta d \left(\frac{1}{2} n \Delta d \Delta \mu -   \Delta \phi \right)\right)}\right|\\
\end{aligned}
\end{equation}
where $\mathbf{a}_t$ and $\mathbf{a}_v$ are any two column of $\mathbf{A}$; $\Delta \mu = \frac{1}{\mu_t} - \frac{1}{\mu_v}$ and $\Delta \phi = \sin\theta_t- \sin\theta_v$. 

The maximum value of $\nu(\mathbf{\Psi})$ is normally obtained when $t$ and $v$ are adjacent grids. We denote the $\Delta \phi_0$ and $\Delta \mu_0$ as the grid interval. A common set of grid intervals is that $\Delta \phi_0 \leq\frac{2}{N}$ and $\Delta \mu_0 \geq\frac{2\lambda}{D^2}$ \cite{9693928, hanChannelEstimationExtremely2020}. In such scenario, we can approximate the second-order function $\frac{1}{2} \Delta d \Delta \mu_0 n^2 - \Delta \phi_0 n$ with a first-order function $ \left(\frac{1}{2} (N-1) \Delta d \Delta \mu_0 -  \Delta \phi_0 \right)n$. Then, we have
\begin{equation}
    \nu(\mathbf{\Psi}) = \sum_{n = 0}^{N - 1}\exp{\left(j k (N-1) \Delta d \left(\frac{1}{2} n \Delta d \Delta \mu_0 -   \Delta \phi_0 \right)\right)}
\end{equation}
It can be observed that $\nu(\mathbf{\Psi}) \simeq 1$ when $\frac{1}{2} (N-1) \Delta d \Delta \mu_0 \simeq \Delta \phi_0$, which is the case that two grids have one distance interval and an angle interval. Hence, $\nu(\mathbf{\Psi})$ with the spatial gridding dictionary $\mathbf{A}$ has big coherence and may not guarantee Lemma \ref{lemma1}.
The proof is completed.
% \cite{xxx}

\ifCLASSOPTIONcaptionsoff
  \newpage
\fi

\bibliographystyle{IEEEtran}
\bibliography{ref}
\end{document}